\documentclass{IEEEtran}
\usepackage{graphicx}
\usepackage{subfigure}
\usepackage[cmex10]{amsmath}
\usepackage{amssymb}
\usepackage{bm}
\usepackage{algorithm}
\usepackage{algorithmic}
\usepackage{cite}
\usepackage{setspace}
\usepackage{stfloats}
\usepackage{multirow}
\usepackage{mathrsfs}
\usepackage{color}
\begin{document}

\makeatletter
\newcommand{\rmnum}[1]{\romannumeral #1}
\newcommand{\Rmnum}[1]{\expandafter \@slowromancap \romannumeral #1@}
\makeatother

\title{Beamforming Design in Multiple-Input-Multiple-Output Symbiotic Radio Backscatter Systems}

\author{
Tuo Wu, Miao Jiang, Qi Zhang, \emph{Member}, \emph{IEEE}, Quanzhong Li, and Jiayin Qin

\thanks{T. Wu, M. Jiang, Q. Zhang, and J. Qin are with the School of Electronics and Information Technology, Sun Yat-sen University, Guangzhou 510006, Guangdong, China (e-mail: wutuo@mail2.sysu.edu.cn, jmiao@mail2.sysu.edu.cn, zhqi26@mail.sysu.edu.cn, issqjy@mail.sysu.edu.cn). Q. Li is with the School of Data and Computer Science, Sun Yat-sen University, Guangzhou 510006, Guangdong, China (e-mail: liquanzh@mail.sysu.edu.cn).}
}

\markboth{}%
{Wu \MakeLowercase{\textit{et al.}}: Beamforming Design in MIMO Symbiotic Radio Backscatter Systems}
\maketitle

\begin{abstract}
Symbiotic radio (SR) backscatter systems are possible techniques for the future low-power wireless communications for Internet of Things devices. In this paper,
we propose a multiple-input-multiple-output (MIMO) SR backscatter system, where
the secondary multi-antenna transmission from the backscatter device (BD) to the receiver is riding on the primary multi-antenna transmission from the transmitter to the receiver. We investigate the beamforming design optimization problem which maximizes the achievable rate of secondary transmission under the achievable rate constraint of primary transmission. In the MIMO SR backscatter system, each antenna of the SR BD reflects its received ambient radio frequency signals from all the transmitting antennas of the transmitter, which causes the globally optimal solution is difficult to obtain. In this paper, we propose a method to obtain the achievable rate upper bound. Furthermore, considering both primary and secondary transmissions, we propose an exact penalty method based locally optimal solution. Simulation results illustrate that our proposed exact penalty method based locally optimal solution performs close to the upper bound.
\end{abstract}


\IEEEpeerreviewmaketitle

\section{Introduction}

Ambient backscatter communication (AmBC), which allows the passive Internet of Things (IoT) devices to transmit their own messages by modulating them over their received ambient radio frequency (RF) signals, achieves high spectrum and energy efficiency \cite{LiuV,HoangDT}. In \cite{WangG} and \cite{QianJ}, for an AmBC system where the backscatter device (BD) adopts differential modulation, the
bit-error-rate (BER) expressions are theoretically derived. In \cite{YangG}, the transceiver design for orthogonal frequency division multiplexing (OFDM) based AmBC system was investigated.

In order to allow the backscattered signals to share the spectrum of primary system, symbiotic radio (SR) backscatter systems were proposed in \cite{KangX,XiaoS,ZhangQ,GuoH,GuoH2}. In \cite{KangX}, Kang \emph{et al.} jointly optimized the transmit power of the primary signal and the reflection coefficient of the secondary AmBC to maximize the ergodic capacity of the secondary system. In \cite{XiaoS}, Xiao \emph{et al.} proposed a full-duplex-enabled SR backscatter system where the throughput of secondary system is maximized under the achievable rate constraints of the primary system. In \cite{ZhangQ}, Zhang \emph{et al.} studied a non-orthogonal multiple access (NOMA) downlink SR backscatter system.
In \cite{GuoH2}, the three relationships between the primary and secondary transmissions, i.e., commensal, parasitic and competitive, in SR backscatter systems were proposed.

Motivated by the benefits of multi-antenna techniques \cite{Andrews,LiQ1,LiQ2}, in this paper, we propose a multiple-input-multiple-output (MIMO) SR backscatter system, where the transmitter, the receiver and the BD are equipped with multiple antennas. Our aim is to optimize the beamforming matrix at the transmitter which maximizes the achievable rate from the BD to the receiver subject to the achievable rate constraint from the transmitter to the receiver. Different from conventional MIMO systems where the channel capacity can be achieved using singular value decomposition (SVD) and water filling, each antenna of the SR BD reflects its received ambient RF signals from all the transmitting antennas of the transmitter, which causes the globally optimal solution to the aforementioned optimization problem is difficult to obtain. In this paper, we propose a method to obtain the achievable rate upper bound. Furthermore, considering both primary and secondary transmissions, we propose an exact penalty method based locally optimal solution.

\emph{Notations}: Boldface lowercase and uppercase letters denote vectors
and matrices, respectively. The conjugate transpose and trace of
the matrix $\mathbf{A}$ are denoted as $\mathbf{A}^H$ and $\mathrm{tr}(\mathbf{A})$, respectively. The $\otimes$ denotes the
Kronecker product. By $\mathbf{A}\succeq\mathbf{0}$, we mean that the matrix $\mathbf{A}$ is positive semidefinite. $\mathcal{CN}(0,\sigma^2)$ denotes the distribution of a circularly symmetric complex Gaussian random variable with zero mean and variance $\sigma^2$.

\section{System Model and Problem Formulation}

Consider a MIMO symbiotic radio backscatter system, as shown in Fig. 1, which consists of a multi-antenna transmitter, a multi-antenna receiver, and a multi-antenna backscatter device (BD). The transmitter, the receiver and the BD are equipped with $N_t$, $N_r$, and $N_b$ antennas, respectively. The channel matrices from transmitter to receiver, from transmitter to BD, and from BD to receiver, are denoted as $\mathbf{G}\in\mathbb{C}^{N_r\times N_t}$, $\mathbf{H}\in\mathbb{C}^{N_b\times N_t}$, and $\mathbf{F}\in\mathbb{C}^{N_r\times N_b}$, respectively.

\begin{figure}
\centering{\includegraphics[width=3in]{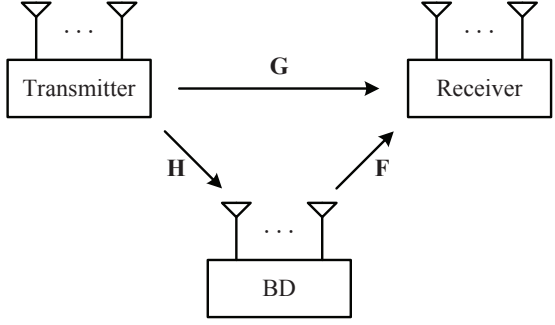}}
\caption{Model of a MIMO symbiotic radio backscatter system which consists of a multi-antenna transmitter, a multi-antenna receiver, and a multi-antenna BD.}
\end{figure}

The signal broadcasted from the transmitter is expressed as
\begin{align}\label{q1}
\mathbf{x}=\mathbf{P}\mathbf{s}
\end{align}
where $\mathbf{s}\in\mathbb{C}^{N_t\times1}$ denotes the signal vector intended to the receiver with $\mathbb{E}{[\mathbf{ss}^{H}]}=\mathbf{I}$ and $\mathbf{P}\in\mathbb{C}^{N_t\times N_t}$ denotes the beamforming matrix.

The BD modulates its own message $\mathbf{c}\in\mathbb{C}^{N_b\times 1}$ with $\mathbb{E}[\mathbf{cc}^{H}]=\mathbf{I}$ intended to the receiver over the received signal vector and backscatters $\mathbf{B}\mathbf{c}$ to the receiver where
\begin{align}\label{q2}
\mathbf{B}=\mathrm{diag}(\mathbf{H}\mathbf{x}).
\end{align}

Thus, the received signal vector at the receiver is
\begin{align}\label{q3}
\mathbf{y}=\mathbf{G}\mathbf{x}+\mathbf{F}\mathbf{B}\mathbf{c}+\mathbf{n}
\end{align}
where $\mathbf{n}\sim\mathcal{CN}(0,\sigma^{2}\mathbf{I})$ denotes the additive Gaussian noise vector received at the receiver. For simplicity, we assume that $\sigma^{2}=1$ throughout this paper. The receiver employs the successive interference cancelation (SIC) to decode the signals from the transmitter and the BD. Since the backscattered signals are generally weak, the BD first decodes the signals from the transmitter and then decodes the signals from the BD. In \eqref{q3}, for the signal decoding from the transmitter, the interference plus noise terms are $\mathbf{F}\mathbf{B}\mathbf{c}+\mathbf{n}$ whose covariance matrix is
\begin{align}\label{q4}
\mathbf{K}=\mathbf{I}+\mathbf{F}\mathbf{D}\mathbf{D}^{H}\mathbf{F}^{H}
\end{align}
where $\mathbf{D}=\mathrm{diag}(\mathbf{H}\mathbf{P}\mathbf{1})$ and $\mathbf{1}$ is an $N_t\times1$ matrix whose entries are all one.

Therefore, the achievable rate from transmitter to receiver is expressed as
\begin{align}\label{q5}
R_t=&\log_{2}\left|\mathbf{I}+\mathbf{K}^{-1}\mathbf{G}\mathbf{PP}^{H}\mathbf{G}^{H}\right|.
\end{align}
If $R_t\geq r_t$ where $r_t$ denotes the achievable rate constraint from transmitter to receiver, the receiver is able to decode $\mathbf{s}$ and remove $\mathbf{G}\mathbf{x}$ in \eqref{q3} by SIC. After SIC, the achievable rate from BD to receiver is
\begin{align}\label{q6}
R_b=\log_{2}\left|\mathbf{K}\right|.
\end{align}

In this paper, our goal is to maximize the achievable rate from BD to receiver
subject to the achievable rate constraint from transmitter to receiver, which is formulated as
\begin{align}\label{q7}
\max_{\mathbf{P}}\ & R_b\nonumber\\
\mbox{s.t.}\ \ & R_t\geq r_t,\nonumber\\
&\mathrm{tr}\left(\mathbf{PP}^H\right)\leq P
\end{align}
where $P$ denotes the transmit power constraint at the transmitter.

\section{Achievable Rate Upper Bound}

In this section, we propose a method to obtain the achievable rate upper bound from BD to receiver subject to the achievable rate constraint from transmitter to receiver. Since
\begin{align}\label{q8}
R_t=&\log_{2}\left|\mathbf{K}+\mathbf{G}\mathbf{PP}^{H}\mathbf{G}^{H}\right|-R_b,
\end{align}
by introducing a slack variable $r_b\geq0$, problem \eqref{q7} is equivalent to
\begin{align}\label{q9}
\max_{\mathbf{P}, r_b\geq0}\ & r_b\nonumber\\
\mbox{s.t.}\quad & \log_{2}\left|\mathbf{K}\right|\geq r_b,\nonumber\\
&\log_{2}\left|\mathbf{K}+\mathbf{G}\mathbf{PP}^{H}\mathbf{G}^{H}\right|\geq r_t+r_b,\nonumber\\
&\mathrm{tr}\left(\mathbf{PP}^H\right)\leq P.
\end{align}
By letting
\begin{equation}
\mathbf{H}=\left[\begin{array}{c}
\mathbf{h}_1^H \\
\vdots \\
\mathbf{h}_{N_b}^H
\end{array}\right],
\end{equation}
we have
\begin{align}\label{q11}
\mathbf{D}=\mathrm{diag}\left(\mathbf{h}_1^H\mathbf{P1},\cdots,\mathbf{h}_{N_b}^H\mathbf{P1}\right).
\end{align}
Therefore,
\begin{align}\label{q12}
\mathbf{D}\mathbf{D}^H=\mathrm{diag}\left(\left|\mathbf{h}_1^H\mathbf{P1}\right|^2,\cdots,\left|\mathbf{h}_{N_b}^H\mathbf{P1}\right|^2\right).
\end{align}
By introducing slack variables
\begin{equation}
\mathbf{Q}=\mathrm{diag}(q_1,\cdots,q_{N_b}),
\end{equation}
problem \eqref{q9} is equivalently transformed into
\begin{subequations}\label{q14}
\begin{align}\label{q14a}
\!\!\!\!\max_{\scriptsize\begin{array}{c}
\mathbf{P},r_b\geq0, \\
\{q_i\geq0\}
\end{array}}& r_b\\
\label{q14b}\mbox{s.t.}\quad\ & \log_{2}\left|\mathbf{I}+\mathbf{F}\mathbf{Q}\mathbf{F}^{H}\right|\geq r_b,\\
\label{q14c}&\log_{2}\left|\mathbf{I}+\mathbf{F}\mathbf{Q}\mathbf{F}^{H}+\mathbf{G}\mathbf{P}\mathbf{P}^H\mathbf{G}^{H}\right|\geq r_t+r_b,\\
\label{q14d}&\left|\mathbf{h}_i^H\mathbf{P1}\right|^2\geq q_i,\ \forall\ i\in\{1,\cdots,N_b\}, \\
\label{q14e}&\mathrm{tr}\left(\mathbf{P}\mathbf{P}^H\right)\leq P.
\end{align}
\end{subequations}
Let $\mathbf{p}=\mathrm{vec}(\mathbf{P})$. We rewrite constraint \eqref{q14d} as
\begin{align}
\mathbf{h}_i^H\mathbf{P1}\mathbf{1}^H\mathbf{P}\mathbf{h}_i\geq q_i,
\end{align}
which is equivalent to
\begin{align}\label{q16}
\mathrm{tr}\left(\mathbf{P1}\mathbf{1}^H\mathbf{P}\mathbf{h}_i\mathbf{h}_i^H\right)\geq q_i.
\end{align}
Using the equality of
\begin{align}
\mathrm{vec}\left(\mathbf{A}_1\mathbf{A}_2\mathbf{A}_3\right)=\left(\mathbf{A}_3^H\otimes\mathbf{A}_1\right)\mathrm{vec}\left(\mathbf{A}_2\right)
\end{align}
where $\mathbf{A}_1$, $\mathbf{A}_2$, and $\mathbf{A}_3$ are arbitrary matrices with compatible dimensions, constraint \eqref{q16} can be transformed into
\begin{align}\label{q18}
\mathrm{tr}\left(\mathbf{p}\mathbf{p}^H\mathbf{H}_i\right)\geq q_i
\end{align}
where $\mathbf{H}_i=\mathbf{h}_i\mathbf{h}_i^H\otimes\mathbf{11}^H$. In constraint \eqref{q14c}, we have
\begin{align}\label{q19}
\mathbf{G}\mathbf{P}\mathbf{P}^H\mathbf{G}^{H}=\sum_{i=1}^{N_t}\mathbf{E}_i\tilde{\mathbf{G}}\mathbf{p}\mathbf{p}^H\tilde{\mathbf{G}}^H\mathbf{E}_i^H
\end{align}
where $\tilde{\mathbf{G}}=\mathbf{I}\otimes\mathbf{G}$ and
\begin{equation}
\mathbf{E}_i=[\mathbf{0}_{N_r\times (i-1)N_r}\ \mathbf{I}_{N_r\times N_r}\ \mathbf{0}_{N_r\times (N_t-i)N_r}].
\end{equation}
Let $\mathbf{\Psi}=\mathbf{p}\mathbf{p}^H$. Employing the rank-one relaxation,  problem \eqref{q14} is recast as
\begin{align}\label{q21}
\!\!\!\!\max_{\scriptsize\begin{array}{c}
\mathbf{\Psi}\succeq\mathbf{0},\\
r_b\geq0, \{q_i\geq0\}
\end{array}}\!\!\!\!\!\!\!\!\!& r_b\nonumber\\
\mbox{s.t.}\quad & \log_{2}\left|\mathbf{I}+\mathbf{F}\mathbf{Q}\mathbf{F}^{H}\right|\geq r_b,\nonumber\\
&\log_{2}\left|\mathbf{I}+\mathbf{F}\mathbf{Q}\mathbf{F}^{H}+\sum_{i=1}^{N_t}\mathbf{E}_i\tilde{\mathbf{G}}\mathbf{\Psi}\tilde{\mathbf{G}}^H\mathbf{E}_i^H\right|\geq r_t+r_b,\nonumber\\
&\mathrm{tr}\left(\mathbf{\Psi}\mathbf{H}_i\right)\geq q_i,\ \forall\ i\in\{1,\cdots,N_b\}, \nonumber\\
&\mathrm{tr}\left(\mathbf{\Psi}\right)\leq P.
\end{align}
It is noted that if the obtained $\mathbf{\Psi}$ is rank-one, it is also the globally optimal solution to problem \eqref{q7}. However, the obtained $\mathbf{\Psi}$ generally has the rank of more than 1. Therefore, the obtained optimal objective value of \eqref{q21} is actually the achievable rate upper bound from BD to receiver subject to the achievable rate constraint from transmitter to receiver.

\section{Exact Penalty Method Based Locally Optimal Solution}

Since the globally optimal solution to problem \eqref{q7} is difficult to obtain, we propose an exact penalty method based locally optimal solution in this section.
By letting $\mathbf{M}=\mathbf{PP}^H$, problem \eqref{q14} is equivalently transformed into
\begin{subequations}\label{qb1}
\begin{align}\label{qb1a}
\!\!\!\!\max_{\scriptsize\begin{array}{c}
\mathbf{M}\succeq0, \mathbf{P}, \\
r_b\geq0, \{q_i\geq0\}
\end{array}}\!\!\!\!\!\!& r_b\\
\label{qb1b}\mbox{s.t.}\quad\ & \log_{2}\left|\mathbf{I}+\mathbf{F}\mathbf{Q}\mathbf{F}^{H}\right|\geq r_b,\\
\label{qb1c}&\log_{2}\left|\mathbf{I}+\mathbf{F}\mathbf{Q}\mathbf{F}^{H}+\mathbf{G}\mathbf{M}\mathbf{G}^{H}\right|\geq r_t+r_b,\\
\label{qb1d}&\left|\mathbf{h}_i^H\mathbf{P1}\right|^2\geq q_i,\ \forall\ i\in\{1,\cdots,N_b\}, \\
\label{qb1e}&\mathrm{tr}\left(\mathbf{M}\right)\leq P,\ \mathbf{M}=\mathbf{PP}^H.
\end{align}
\end{subequations}
To continue, we have the following lemma whose proof can be found in \cite{Rashid}.

\emph{Lemma 1}: The equality constraint $\mathbf{M}=\mathbf{PP}^H$ in problem \eqref{qb1} is equivalent to
\begin{align}\label{qb2}
\left[
  \begin{array}{ccc}
    \mathbf{W}_{1} & \mathbf{M} & \mathbf{P} \\
    \mathbf{M}^{H} & \mathbf{W}_{2} & \mathbf{P} \\
    \mathbf{P}^{H} & \mathbf{P}^{H} & \mathbf{I} \\
  \end{array}
\right]\succeq 0, \\
\label{qb3}\mathrm{tr}(\mathbf{W}_1-\mathbf{P}\mathbf{P}^{H})\leq 0,
\end{align}
where $\mathbf{W}_{1}\in\mathbb{C}^{N_t\times N_t}$ and $\mathbf{W}_{2}\in\mathbb{C}^{N_t\times N_t}$ are slack variables.

$\hfill\blacksquare$

Replacing the constraint $\mathbf{M}=\mathbf{PP}^H$ with \eqref{qb2} and \eqref{qb3}, problem \eqref{qb1} is still non-convex because of the non-convex constraints \eqref{qb1d} and \eqref{qb3}. To solve problem \eqref{qb1}, we define a compact and convex set
\begin{align}\label{qb4}
\Omega=\{\chi=&\left.\left(\mathbf{M}, \mathbf{P}, r_b, \{q_i\}, \mathbf{W}_1,\mathbf{W}_2\right)\right|\eqref{qb1b},\eqref{qb1c},\eqref{qb2}, \nonumber\\
&\mathbf{M}\succeq0, r_b\geq0, q_i\geq0, \mathrm{tr}\left(\mathbf{M}\right)\leq P\}.
\end{align}
Using Lemma 1, we equivalently rewrite the problem \eqref{qb1} as
\begin{align}\label{qb5}
&\max_{\chi\in\Omega}\ r_b\ \ \mbox{s.t.}\ \eqref{qb1d},\ \eqref{qb3}.
\end{align}

To solve problem \eqref{qb5}, we employ the exact penalty method \cite{LiQ1,Rashid} and rewrite it as
\begin{align}\label{qb6}
&\min_{\chi\in\Omega}\ \mathcal{L}\ \ \mbox{s.t.}\ \xi_i\geq q_i,\ \forall\ i\in\{1,\cdots,N_b\}
\end{align}
where
\begin{align}\label{qb7}
\mathcal{L}=&-r_b+\mu\mathrm{tr}\left(\mathbf{W}_1-\mathbf{P}\mathbf{P}^H\right),\\
\label{qb8}
\xi_i=&\left|\mathbf{h}_i^H\mathbf{P1}\right|^2,
\end{align}
and $\mu$ denotes the Lagrangian dual variable.

To proceed, we have the following lemma whose proof can be found in \cite{LiQ1,Rashid}.

\emph{Lemma 2}: There exists $ 0<\mu^\star<+\infty $ such that problems \eqref{qb5} and \eqref{qb6} are equivalent when $\mu>\mu^\star$.
$\hfill\blacksquare$

Rewrite $\mathcal{L}$ as
\begin{align}
\mathcal{L}=-r_b+\mu\mathrm{tr}\left(\mathbf{W}_1\right)-\zeta
\end{align}
where
\begin{align}\label{qb10}
\zeta=\mu\mathrm{tr}\left(\mathbf{P}\mathbf{P}^H\right).
\end{align}
From \eqref{qb8} and \eqref{qb10}, problem \eqref{qb6} is a difference of convex (DC) programming because both $\xi_i$ and $\zeta$ are convex. We propose to employ the constrained concave convex procedure (CCCP) to solve problem \eqref{qb6}. The first-order Taylor expansions of $\xi_i$ and $\zeta$ around the point $\tilde{\mathbf{P}}$ are
\begin{align}
\xi_i&\geq\left|\mathbf{h}_i^H\tilde{\mathbf{P}}\mathbf{1}\right|^2+2\mathrm{Re}\left[\mathbf{h}_i^H\tilde{\mathbf{P}}\mathbf{11}^H\left(\mathbf{P}-\tilde{\mathbf{P}}\right)\mathbf{h}_i\right],\\
\zeta&\geq\mu\mathrm{tr}\left(\tilde{\mathbf{P}}\tilde{\mathbf{P}}^H\right)+2\mu\mathrm{Re}\left\{\mathrm{tr}\left[\left(\mathbf{P}-\tilde{\mathbf{P}}\right)\tilde{\mathbf{P}}^H\right]\right\}.
\end{align}

Thus, in the $(l+1)$th iteration, given $\mathbf{P}_l$ which is optimal in the $l$th iteration, we solve the following convex optimization problem to obtain $\mathbf{P}_{l+1}$
\begin{align}
\min_{\chi\in\Omega}\ &-r_b+\mu\mathrm{tr}\left(\mathbf{W}_1\right)-\mu\mathrm{tr}\left(\mathbf{P}_l\mathbf{P}_l^H\right)\nonumber\\
&+2\mu\mathrm{Re}\left\{\mathrm{tr}\left[\left(\mathbf{P}-\mathbf{P}_l\right)\mathbf{P}_l^H\right]\right\}\nonumber\\
\mbox{s.t.}\ &\left|\mathbf{h}_i^H\mathbf{P}_l\mathbf{1}\right|^2+2\mathrm{Re}\left[\mathbf{h}_i^H\mathbf{P}_l\mathbf{11}^H\left(\mathbf{P}-\mathbf{P}_l\right)\mathbf{h}_i\right]\geq q_i,\nonumber\\
&\quad\quad\forall\ i\in\{1,\cdots,N_b\}.
\end{align}
When it converges, we obtain the locally optimal solution to problem \eqref{q7}.

\section{Simulation Results}

In this section, we present the computer simulation results to validate our proposed beamforming designs. In simulations, we assume that the transmitter, the receiver and the BD are equipped with $N_t=N_r=N_b=2$ antennas. The channels are independent and identically distributed (i.i.d.) block flat Rayleigh fading channels such that $\mathbf{G}\sim\mathcal{CN}(\mathbf{0},\mathbf{I})$, $\mathbf{H}\sim\mathcal{CN}(\mathbf{0},\mathbf{I})$, and $\mathbf{F}\sim\mathcal{CN}(\mathbf{0},\mathbf{I})$. The achievable rate constraint from the transmitter to the receiver is $r_t=2$ bps/Hz.

In Fig. 2, we present the convergence behavior of our proposed exact penalty method based locally optimal solution for different values of $P/\sigma^2$. From Fig. 2, it is observed that our proposed exact penalty method based locally optimal solution converges for about 3 iterations.

\begin{figure}
\centering{\includegraphics[width=3.6in]{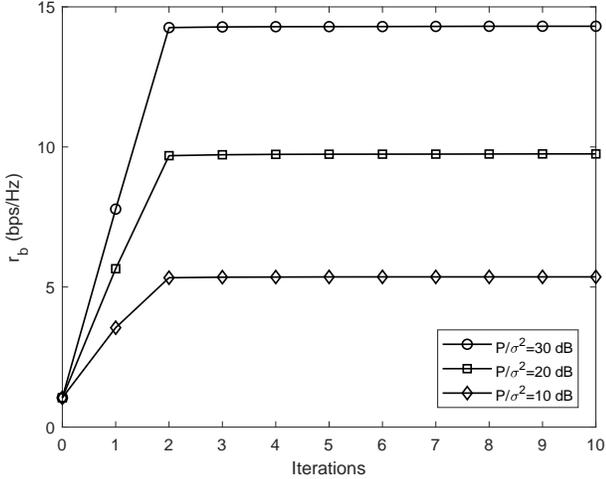}}
\caption{Convergence behavior of our proposed exact penalty method based locally optimal solution for different values of $P/\sigma^2$.}
\end{figure}

In Fig. 3, we present the achievable rate $r_b$ comparison of our proposed exact penalty method based locally optimal solution with the achievable rate upper bound, denoted as ``EPM-LO" and ``UB" in the legend, respectively. In Fig. 3, we also present the achievable rate $r_b$ of the maximum ratio transmission (MRT) schemes. For MRT schemes, we mean that using singular value decomposition (SVD), the channel matrices $\mathbf{G}$ and $\mathbf{H}$ can be expressed as
\begin{align}
\mathbf{G}=\mathbf{U}_g\bm{\Lambda}_g\mathbf{V}_g,\\
\mathbf{H}=\mathbf{U}_h\bm{\Lambda}_h\mathbf{V}_h,
\end{align}
where $\mathbf{U}_g$, $\mathbf{U}_h$, $\mathbf{V}_g$, and $\mathbf{V}_h$ are unitary matrix; $\bm{\Lambda}_g$ and $\bm{\Lambda}_h$ are diagonal matrices. The beamforming matrices are the weighted $\mathbf{V}_g$ and $\mathbf{V}_h$, denoted as ``MRT-G" and ``MRT-H" in the legend, respectively. From Fig. 3, it is found that the achievable rate $r_b$ obtained by our proposed exact penalty method based locally optimal solution is close to the achievable rate upper bound and is higher
than both ``MRT-G" and ``MRT-H" schemes.

\begin{figure}
\centering{\includegraphics[width=3.6in]{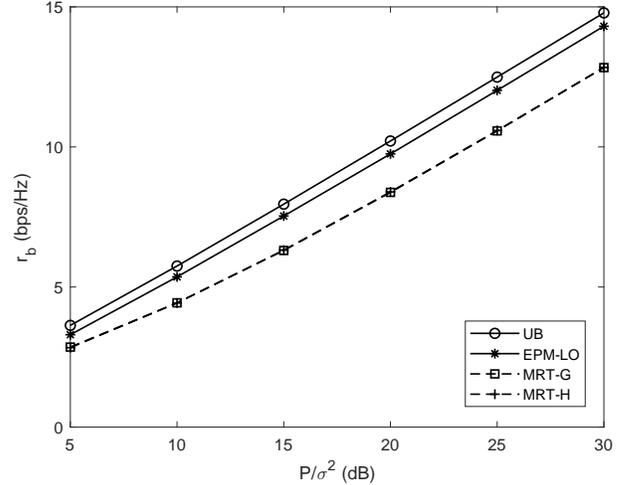}}
\caption{Achievable rate $r_b$ versus $P/\sigma^2$; comparison of our proposed exact penalty method based locally optimal solution with the achievable rate upper bound, the ``MRT-G" and ``MRT-H" schemes. }
\end{figure}

\section{Conclusion}

In this paper, we have studied a MIMO SR backscatter system, where the secondary multi-antenna transmission from BD to receiver is riding on the primary multi-antenna transmission from transmitter to receiver. For the MIMO SR backscatter system, we have proposed a method to obtain the achievable rate upper bound. Furthermore, considering both primary and secondary transmissions, we have proposed an exact penalty method based locally optimal solution. It is shown through simulation results that our proposed exact penalty method based locally optimal solution obtains the achievable rate which is slightly lower than the upper bound and much higher than the MRT schemes.

\end{document}